\documentclass[prb,aps]{revtex4}
\usepackage{graphics}
\begin{document}
 
\title{Modified spin-wave study of random
antiferromagnetic-ferromagnetic spin chains}

\author{Xin Wan and Kun Yang}
\affiliation{National High Magnetic Field Laboratory and Department of Physics,
Florida State University, Tallahassee, Florida 32306}

\author{R. N. Bhatt}
\affiliation{Department of Electrical Engineering, Princeton University,
        Princeton, New Jersey 08544}

\date{\today}

\begin{abstract}
We study the thermodynamics of one-dimensional quantum spin-1/2
Heisenberg ferromagnetic system with random antiferromagnetic impurity
bonds. 
In the dilute impurity limit, we generalize the modified spin-wave
theory for random spin chains, where local chemical potentials for
spin-waves in ferromagnetic spin segments are introduced to ensure zero
magnetization at finite temperature.
This approach successfully describes the crossover from behavior of
pure one-dimensional
ferromagnet at high temperatures to a distinct Curie behavior due to
randomness at low temperatures. 
We discuss the effects of impurity bond strength and concentration on the
crossover and low temperature behavior.  
\end{abstract}


\maketitle

\section{Introduction}
The renormalization group (RG) approach of Wilson~\cite{wilson74}
provided a framework in which to understand the behavior of various phases
of thermodynamic systems in terms of trajectories and RG
flows in Hamiltonian space. In addition, it provided a scheme,
to calculate, in principle, the behavior in the vicinity of
critical and fixed points, by linearizing the flow equations.
The method has proved to be powerful in providing an
understanding of uniform (translationally invariant) systems,
an exact implementation of the Wilson method in systems with
quenched random disorder has been relatively restricted.
Thus, for example, while the method has been successfully
applied to the Anderson localization problem in ($ 2 + \epsilon $)
dimensions,~\cite{wegner76} generalization to include electron-electron
interactions~\cite{belitz94}
have had limited success, because of the complications
such as the existence of local moments~\cite{milovanovic89,bhatt92} 
which are left out in such long-wavelength
approaches to the problem.
Another class of problems where a large amount of work
has been done~\cite{dhbb}
is based on the random antiferromagnetic (AF) spin-1/2 chain in one
dimension,
where it has been shown~\cite{fisher94} that
a perturbative real space renormalization group (RSRG) scheme
becomes asymptotically exact, as has been explicitly verified
in a number of cases, such as the random XY chain, where it can be
mapped on to a model of free fermions.~\cite{mckenzie96}

Another issue where less attention has been given, is the range of
validity
of the asymptotic results of the linearized RG equations. In most models
of uniform
systems, a few iterations of the RG equations leads to the vicinity of
the appropriate fixed point. Consequently, a study of the RG trajectory
from the region in the vicinity of an critical fixed point to a
stable fixed point, usually yields a relatively quick crossover
behavior.
However, in a study of a three dimensional model related to the 1D
random AF
chain,~\cite{bhatt82} it was found that the system
did not settle down to a fixed-point-like behavior over several orders
of magnitude
in temperature; instead there was a slow, logarithmic in temperature,
evolution
in the behavior. Indeed, it has been claimed recently~\cite{motrunich00}
that the true low temperature behavior of the model is likely different.
Unfortunately,
such a change cannot be experimentally investigated, because smaller
terms in the
full Hamiltonian (such as hyperfine coupling to nuclear spins), and
dipolar couplings,
which have been neglected in the idealized model, will modify the
behavior at such
(exponentially low) temperatures. Nevertheless,
it raises the issue that in models with quenched disorder, the true
asymptotic
fixed point behavior, may not be practically reached, at least in some
cases,
making ``relevant" (in the RG sense) irrelevant in any practical sense.
It is therefore of considerable interest to examine models with quenched
disorder where
crossovers between short length scale and long length scale behavior can
be
studied in some detail.

One model, which has received extensive attention in recent years, is
the random antiferromagnetic-ferromagnetic (AFM-FM) spin 
chain.~\cite{furusaki94,westerberg95,furusaki95,westerberg97,hida97,frischmuth97,frischmuth99,hikihara99}
The study of such systems has been motivated by the discovery of novel
one-dimensional (1D) random spin systems. 
One example of such materials is
Sr$_3$CuPt$_{1-x}$Ir$_x$O$_6$,~\cite{nguyen96} 
alloy of the pure compounds 
Sr$_3$CuPtO$_6$ (antiferromagnet) and Sr$_3$CuIrO$_6$ (ferromagnet).
Such a system is interesting due to the interplay between quantum fluctuations
and disorder; the latter can not be treated as a perturbation, 
as it is often found to change the low-energy spectra dramatically or 
even to destabilize the pure phases. 
Theoretical and numerical methods used to study random
antiferromagnetic-ferromagnetic spin chains have been largely inherited
from the study of random antiferromagnetic (AFM) 
chains (with AFM coupling only).~\cite{ma79,fisher94,hyman96} 
In particular, the real-space renormalization group (RSRG) 
approach~\cite{westerberg95,westerberg97} revealed a
Curie-like temperature dependence in the susceptibility of 
a random AFM-FM spin-1/2 chain at low temperatures.
In the RSRG picture, the Curie behavior occurs because
spins correlate, due to the existence of
ferromagnetic couplings, to form clusters whose average size and
effective spin grow in a random walk fashion at low temperatures.
Therefore, the random AFM-FM spin-1/2 chain belongs to a different
universality class from that of the random AFM spin-1/2 chain, 
whose ground state is known as a random singlet phase, 
in which singlets can be formed over large distance. 
The RSRG results have been supported by various other numerical
techniques, such as quantum Monte Carlo (QMC) 
simulation~\cite{frischmuth97,frischmuth99,ammon99} and density
matrix renormalization group (DMRG) method.~\cite{hida97,hikihara99} 

In the RSRG method, one decimates the spins that are coupled by the strongest
coupling in the system, and renormalizes the couplings among the remaining spins
perturbatively. In this procedure new effective
couplings and effective spins may also be
generated. This procedure, which in general lowers the overall energy scale of 
the problem at hand, is repeated so that excitations at lower and lower energy
scales are probed. Its quantitative accuracy relies on the presence of 
{\em strong} randomness, as in the case of strong randomness the strength 
of the strongest and typical couplings are well separated, ensuring
the accuracy of a perturbative calculation. In real systems, however, the
randomness can be very weak, especially when it is not introduced intentionally.
In this case the RSRG is {\em not} expected to be quantitatively reliable in its
early stages. It has been argued\cite{fisher94,westerberg97}
that the strength of disorder grows 
as the energy scale is lowered, thus in the low-energy (or equivalently, 
low-temperature) limit, the RSRG results become correct or even asymptotically
exact. Nevertheless in this case one expects the thermodynamic properties of the
system to be dominated by the physics of the pure chains at high temperature,
and crossover to the randomness dominated regime (where the RSRG results
apply) at very low temperatures.

In the present work we study the thermodynamic properties of the random
AFM-FM chain, with mostly uniform FM coupling and a small concentration of 
(impurity) AFM bonds, using the modified spin-wave method. We also compare 
the results of modified spin-wave method with those of exact diagonalization
in small systems to demonstrate the accuracy of the former.
Our motivation comes from the following considerations.
First of all, as discussed above, when the concentration of the impurity bonds
is small, the randomness is rather weak and the RSRG results 
are not reliable at high temperatures. The modified spin-wave method allows us
to treat the high and low temperature regimes on equal footing, and in 
particular, address the crossover between these two regimes. This makes it 
possible to compare theory with experiments for the entire temperature range, 
when the randomness is weak. Secondly, the modified spin-wave method has so 
far been applied to study pure spin models only. While very successful in those
cases, it has not yet been used to study random spin systems. By working out
a number of technical issues that one faces when applying it to random spin 
problems, and demonstrate its accuracy in the present problem, we lay the 
ground for the application of this powerful method to other random spin 
problems. Thirdly, as discussed in the opening paragraphs, it is of
general interest to study models where the short distance (high
temperature) and long distance (low temperature) behavior are controlled
by different fixed points in the renormalization group sense, and address
the crossover between the two limiting cases
quantitatively; our model is such an example, and
we hope our study will stimulate
future research on this important issue. 

Our results are summarized as follows. We find that the spin susceptibility of
our system follows that of a pure ferromagnetic chain at high temperatures, and
crosses over to $1/T$ dependence at low temperatures predicted by the RSRG, 
with a coefficient that agrees with the RSRG result essentially exactly. We also
find that the crossover temperature depends mainly on the impurity bond 
{\em concentration}, while the width of the crossover depends mainly on the 
impurity bond
{\em strength}; we determine their dependences semiquantitatively. 
We also demonstrate 
that the results of the modified spin-wave method is asymptotically exact 
in certain limiting cases, and quite accurate in the entire temperature range by
comparing them with exact diagonalization results in the systems.

The paper is organized as follows. 
In Sec.~\ref{sec:picture}, we describe the model we study, as well as
the qualitative physics expected from the model.
In Sec.~\ref{sec:method}, we introduce the modified spin-wave theory for
the random AFM-FM spin chains.  
We present the numerical results of the theory in
Sec.~\ref{sec:results}, with emphasis on the effects
impurity bond strength, impurity concentration, as well as finite-size effects 
of our numerical study on 
the spin susceptibility. 
Finally, our results are summarized in Sec.~\ref{sec:conclusion}. 
Readers interested in the comparison of the modified spin-wave theory
to exact diagonalization results will find details in 
Appendix~\ref{singleSegment}
and \ref{twoSegments} for simple cases like single ferromagnetic segment
and two coupled segments. 

\section{The qualitative picture}
\label{sec:picture}

We consider the following Hamiltonian for a random spin-$S$ chain
\begin{equation}
\label{hamiltonian}
H = \sum_i J_i {\bf S}_i \cdot {\bf S}_{i+1},
\end{equation}
where the coupling strength $J_i$ is randomly chosen according to the
distribution 
\begin{equation}
P(J) = (1 - p) \delta (J + J_F) + p \delta (J - J_{AF}).
\end{equation}
$J_F$ and $J_{AF}$ are the coupling strengths of the ferromagnetic and
antiferromagnetic bonds, respectively. 
$p$ is the concentration of the antiferromagnetic impurity bonds. 
We are primarily interested in the dilute doping regime, 
{\it i.e.}, $p \ll 1$, so that we have long ferromagnetic spin segments,
separated by antiferromagnetic bonds.  
The number of spins in a spin segment ($N_s$) is random, 
with an exponential distribution 
\begin{equation}
\label{segmentdist}
P(N_s) =  p (1-p)^{N_s-1}.
\end{equation}
Therefore, the average number of spins per segment 
is $\langle N_s \rangle = 1 / p$. 
From real space renormalization group studies\cite{ma79,bhatt82}, 
we understand that a strong antiferromagnetic bond tends to lock the two
adjacent spins into a singlet (which is inert at temperatures of
interest), and the singlet thereby generates a weaker effective AF coupling 
between the two spins next to the singlet (see
Appendix~\ref{twoSegments}).  
Therefore, for simplicity, we consider weak antiferromagnetic coupling
($J_{AF} < J_F$) only without losing generality. 

The temperature dependence of magnetic susceptibility $\chi$ per spin 
is qualitatively known. 
We assume the Zeeman term $H' = -\mu H_z \sum S_i^z$ and 
set $\mu = k_B = 1$ for simplicity.
At high temperatures ($T \gg J_F$), the spins behave independently. 
Therefore, we expect an ordinary Curie contribution of 
\begin{equation}
\label{highTexpansion}
\chi_{HT} = {S (S+1) \over 3T},
\end{equation}
from each spin. 
When $T \leq J_F$, the spins within each ferromagnetic segment become
correlated.
Takahashi~\cite{takahashi85} obtained the following low temperature
expansion for spin-1/2 ferromagnetic chain (of infinite length)
\begin{equation}
\label{lowTexpansion}
\chi_{FM} = {1 \over 4} \left [ {0.1667 J_F \over T^2} 
+ {0.581 J_F^{1/2} \over T^{3/2}} + {0.68 \over T} 
+ O \left ( T^{1/2} \right ) \right ]
\end{equation}
using the Bethe-ansatz integral equations. 
The factor of $1/4$ is added here due to an extra factor of 2 in the
Zeeman energy terms in the original paper.
Therefore, susceptibility of each segment rises as temperature decreases, 
crossing over from $1 / T$ to $1 / T^2$. 
In the case of $J_{AF} = 0$, $\chi$ crosses over back to Curie behavior 
at lower temperatures, when each segment of finite length acts like 
a block spin with frozen internal excitations.  
When $T$ drops below the typical spin-wave excitation gap of a typical 
segment
\begin{equation}
\Delta_{SW} \sim J_F S \left ( {2 \pi \over \bar{N}_s} \right )^2,
\end{equation}
we expect 
\begin{equation}
\chi_{BS} \equiv {c_{BS} \over T} = {1 \over N} 
\sum_s {N_s S (N_s S + 1) \over 3T}.
\end{equation}
Here $\bar{N}_s$ is the typical length of a ferromagnetic segment.
In the thermodynamic limit, we obtain the Curie constant for
locked block spins to be
\begin{equation}
\label{independentSegments}
c_{BS} = {S^2 \over 3} {2 - p \over p} + {S \over 3},
\end{equation}
using segment-size distribution in Eq.~\ref{segmentdist}.
At $p = 0.1$, we obtain $c_{BS} = 7/4$ for $S = 1/2$. 
$c_{BS}$ decreases with increasing impurity concentration $p$ and is 
always (unless $p = 1$, {\it i.e.} in the pure antiferromagnetic case)
greater than the high temperature Curie constant (in Eq.~\ref{highTexpansion})
\begin{equation}
\label{cht}
c_{HT} = {S (S+1) \over 3} = {1 \over 4}, 
{\rm \ for \ } S = {1 \over 2},
\end{equation}
expected at high temperatures. 

When the end spins of neighboring segments are coupled with weak
antiferromagnetic coupling $J_{AF}$, the block spins (originating from
the ferromagnetic segments) interact with weak effective couplings. 
For two segments with $N_1$ and $N_2$ spins, we obtain the effective
coupling 
\begin{equation}
\label{effectiveJ}
J' = {J_{AF} \over N_1 N_2},
\end{equation}
by projecting the coupling to the block-spin space using Wigner-Eckart
theorem.  
Not surprisingly, long segments are weakly coupled. 
In the RSRG approach,\cite{furusaki95} one starts with two strongly
coupled segments and decimates the segments to a $S' = |N_1 - N_2| S$
spin, and renormalize its couplings with neighboring segments.
The spin-wave excitations within each segment, considered in our
modified spin-wave approach, are in general neglected in the RSRG
approach, which is, therefore, valid only at low enough temperatures.  
By including these spin-wave excitations, we can not only demonstrate 
that the Curie susceptibility (predicted by RSRG) 
in a AFM-FM spin-1/2 chain at low temperatures is indeed  
the result of correlated ferromagnetic segments, but estimate the
temperature scale at which the low temperature Curie behavior occurs as
well. 
Since the renormalized couplings may become ferromagnetic, 
clusters start to grow in a random walk fashion 
as segments couple to neighboring segments as we lower the temperature. 
As in a typical random walk problem, we expect the average spin
$\bar{S}_l$ and the average size $\bar{l}$ of such clusters of spins 
satisfy
\begin{equation}
\bar{S}_l \sim \bar{l}^{-1/2},
\end{equation}
so that as far as magnetic susceptibility per spin $\chi$ is concerned, 
the scaling behavior of $\bar{S}_l^2$ and $\bar{l}$ cancels, leading to
\begin{equation}
\chi_{LT} = {c_{LT} \over T} = 
{1 \over 3T} {\bar{S}_l^2 \over \bar{l}} \sim {1 \over T}.
\end{equation}
By statistical analysis,\cite{westerberg97} one finds the
low-temperature Curie constant to be 
\begin{equation}
\label{rsrg}
c_{LT} = {S^2 \over 3} {1 - p \over p},
\end{equation}
in the large cluster-size limit.
This result coincides with the classical result for the susceptibility
per spin at low temperatures.~\cite{furusaki94,nguyen96}  
Compared with Eq.~\ref{independentSegments}, we find that
antiferromagnetic impurity bonds reduce the low-temperature Curie constant
by  
\begin{equation}
\Delta c = c_{BS} - c_{LT} = {S^2 \over 3p} + {S \over 3}.
\end{equation}
On the other hand, for small $p$ (or low concentration of AF bonds), we have
$c_{LT} > c_{HT}$; this simply reflects the fact that the dominant interaction
in the system is ferromagnetic, which enhances spin susceptibility. Thus in the
cases we are interested in, we have
\begin{equation}
c_{HT} < c_{LT} < c_{BS}.
\end{equation}

We should note here that unless $J_{AF}$ is very small, there may not be a 
temperature range in which $\chi\approx \chi_{BS}= c_{BS}/T$; this is 
because the neighboring segments can start developing correlations due to 
$J_{AF}$ {\em before} the intrasegment spin-wave excitations are frozen out by
temperature. We will actually study mostly this case in this work, and focus
mainly on the crossover from the pure ferromagnetic chain behavior directly to
the asymptotic low temperature behavior with $c_{LT}$ as the Curie coefficient.
We do this partly because in real systems $J_{AF}$ is often comparable to 
$J_F$, and partly for the sake of simplicity. Our method, however, is  
capable of handling very small $J_{AF}$ and the regime with $c_{BS}$ being
the Curie coefficient.

\section{The modified spin-wave approach}
\label{sec:method}

The conventional spin-wave theory gives the exact spin-wave spectra of
Heisenberg ferromagnets at $T=0$. At finite temperature however,
the theory leads to difficulty in
one-dimensional quantum ferromagnets, 
the number of spin waves diverges
when external magnetic field goes to zero. 
This difficulty has its root in the Mermin-Wagner theorem, which dictates that
there cannot be long-range order in 1D at finite temperature.
By imposing a constraint that the total magnetization be zero (thus fulfilling
the Mermin-Wagner theorem),
Takahashi\cite{takahashi86} introduced a chemical potential (equivalence
to a uniform magnetic field) for the spin waves and successfully
obtained the low-temperature properties of the ferromagnets in one
dimension, as well as in two dimensions.  
The beauty of this modified spin-wave theory is that it gives correct
asymptotic thermodynamic behavior at both large and zero temperature
limit, as well as fairly accurate crossover behavior at intermediate
temperatures. 
The modified spin-wave theory have further succeeded in 
two-dimensional antiferromagnets\cite{takahashi89,tang89} and
one-dimensional ferrimagnets.\cite{yamamoto98}
For one-dimensional antiferromagnets, the modified spin-wave theory
predicted a gap in the spin-wave spectrum,\cite{rezende90} consistent
with the Haldane gap in integer-spin chains.\cite{haldane83}

In this paper, we extend its applicability to ferromagnetic spin chains
with dilute antiferromagnetic impurities. 
In the dilute impurity limit, a spin chain consists of 
ferromagnetic spin segments, coupled antiferromagnetically.  
The application of the modified spin-wave theory to the random
antiferromagnetic-ferromagnetic spin chains is, therefore, mainly based
on our observation that ferromagnetic spin segments can be well
described by the modified spin-wave theory (see
Appendix~\ref{singleSegment}). 
We found that, for magnetic susceptibility of a finite spin chain, the
modified spin-wave theory gives noticeable difference from the exact
diagonalization result only at two temperature crossovers.
The crossover at higher temperatures describes the switching on of the
couplings between individual spins, which is of little interest in the
competition between different segments. 
On the other hand, the maximum error of the modified spin-wave theory is
roughly 5\%, for a chain of up to 14 spins, near the lower-temperature
crossover when the spin chain becomes locked into one block spin.
For a finite spin chain, the translational symmetry is broken. 
Therefore, one, in principle, needs to introduce a local chemical
potential (or a local field) for each spin to ensure the magnetic moment
of the spin be zero. 
Fortunately, we can apply a periodic boundary condition to each
ferromagnetic spin segment, so that only one chemical potential is
needed throughout the segment. 
We emphasize that the use of periodic boundary condition reduces the
computational complexity without changing the essential physics.
This is particularly true when we have very dilute impurities so that
the average ferromagnetic segment length is large. 

For the nearest-neighbor Heisenberg Hamiltonian, Eq.~\ref{hamiltonian}, 
we apply Holstein-Primakoff
transformation
\begin{equation}
\left \{
\begin{array}{lll}
S_i^{s+} &=& \sqrt{2S - a_i^{s\dagger} a_i^s} a_i^s, \\
S_i^{s-} &=& a_i^{s\dagger} \sqrt{2S - a_i^{s\dagger} a_i^s}, \\
S_i^{sz} &=& S - a_i^{s\dagger} a_i^s. 
\end{array}
\right .,
\end{equation}
{\it for segments with odd index $s$}, and
\begin{equation}
\left \{
\begin{array}{lll}
S_i^{s+} &=& a_i^{s\dagger} \sqrt{2S - a_i^{s\dagger} a_i^s}, \\
S_i^{s-} &=& \sqrt{2S - a_i^{s\dagger} a_i^s} a_i^s, \\
S_i^{sz} &=& a_i^{s\dagger} a_i^s - S. 
\end{array}
\right .,
\end{equation}
{\it for segments with even index $s$}.
This way, we introduce a distinct species of boson for each segment. 
The Hamiltonian, in the linear spin-wave approximation, becomes
\begin{eqnarray}
H = E_0 
&+& J_F S \sum_{s = 1}^{M_s} \sum_{i = 1}^{N_s} 
\left ( a_i^{s\dagger} a_i^s 
+ a_{i+1}^{s\dagger} a_{i+1}^s 
- a_i^{s\dagger} a_{i+1}^s 
- a_{i+1}^{s\dagger} a_i^s \right ) 
\nonumber \\ 
&+& J_{AF} S \sum_{s = 1}^{M_s-1} 
\left ( a_{N_s-1}^{s\dagger} a_{N_s-1}^s 
+ a_1^{(s+1)\dagger} a_1^{(s+1)} 
+ a_{N_s-1}^{s\dagger} a_1^{(s+1)} 
+ a_1^{(s+1)\dagger} a_{N_s-1}^s \right ),
\end{eqnarray}
where $M_s$ is the number of ferromagnetic segments, and $N_s$ the
number of spins in the $s$-th segment. 
The total number of spins is, therefore, $N = \sum_{s=1}^{M_s} N_s$.   
Note $(N_s + 1) \equiv 0$, reflecting the periodic boundary condition
imposed on each segment.

In the linear spin-wave approach, the quadratic Hamiltonian is soluble
by a generalized Bogoliubov transformation for bosons, a generalization
of Ref.~\onlinecite{colpa78}. 
The procedure can be written compactly in a matrix format. 
Denote the original boson operators by a vector
\begin{equation}
x = (a_1^1, a_2^1, \cdots, a_{N_1}^1, a_1^{2\dagger}, a_2^{2\dagger}, 
\cdots, a_{N_2}^{2\dagger}, \cdots)^T.
\end{equation}
The Hamiltonian can be written as 
\begin{equation}
H = const. + x^{\dagger} {\cal H} x,
\end{equation}
where
\begin{equation}
\label{hamiltonianMatrix}
{\cal H} = J_F S \left ( 
\begin{array}{cccccccccc}
2 & -1 & & -1 & &  &  &  & \\
-1 & 2 & \ddots  &  &  &  &  &  &  & \\
 & \ddots  & \ddots  & -1 &  &  &  &  &  & \\
-1 &  & -1 & 2+J_{AF}/J_F &  J_{AF}/J_F &  &  &  &  & \\
 &  &  & J_{AF}/J_F & 2+J_{AF}/J_F & -1 &  & -1  &  & \\
 &  &  &  & -1 & 2 & \ddots  &  &  & \\
 &  &  &  &  & \ddots  & \ddots  & -1  &  & \\
 &  &  &  & -1 &  & -1 & 2+J_{AF}/J_F & J_{AF}/J_F  & \\
 &  &  &  &  &  &  & J_{AF}/J_F & 2+J_{AF}/J_F & \ddots \\
 &  &  &  &  &  &  &  & \ddots & \ddots 
\end{array}
\right) ,
\end{equation}
is a $N \times N$ matrix. 
The Hamiltonian matrix is basically a 3-diagonal matrix, with extra
(-1)'s at the corners of each block (of a spin segment). 
The rest of the matrix elements (left unspecified in
Eq.~\ref{hamiltonianMatrix}) are zeros.  
Introduce the generalized Bogoliubov transformation:
\begin{equation}
x = V \gamma,
\end{equation}
where $V$ is a $N \times N$ matrix and $\gamma$ a vector of the set of
boson operators ($\alpha$'s) 
\begin{equation}
\gamma = (\alpha_1^1, \alpha_2^1, \cdots, \alpha_{N_1}^1, 
\alpha_1^{2\dagger}, \alpha_2^{2\dagger}, \cdots, 
\alpha_{N_2}^{2\dagger}, \cdots)^T,
\end{equation}
that diagonalizes ${\cal H}$, {\it i.e.}, 
\begin{equation}
\label{diagonalization}
V^T {\cal H} V = E,
\end{equation}
in the compact matrix format, where $E$ is a diagonal matrix,
\begin{equation}
E_{N\times N} = diag (\varepsilon_1^1, \varepsilon_2^1, \cdots, 
\varepsilon_{N_1}^1, \varepsilon_1^{2\dagger}, \varepsilon_2^{2\dagger}, 
\cdots, \varepsilon_{N_2}^{2\dagger}, \cdots, 
\varepsilon_{N_{M_s}}^{M_s\dagger})
\end{equation}
with its diagonal elements being the
energies of the corresponding bosons after the Bogoliubov transformation.
The boson commutation relation requires
\begin{equation}
\label{normalization}
V P V^T = P,
\end{equation}
where $P$ is a generalized {\it para unit matrix} (inheriting the
notation from Ref.~\onlinecite{colpa78}) of the form 
\begin{equation}
P_{N\times N}= diag (\overbrace{1, 1, \cdots, 1}^{N_1}, 
\overbrace{-1, -1, \cdots, -1}^{N_2}, \cdots )
\end{equation}
For fermions, on the other hand, the fermionic commutation relations
lead to the orthonormalization condition, which can assume the same
equation as Eq.~\ref{normalization}, except that $P$ becomes an $N \times
N$ unit matrix. 
We need to find out a solution $V$ for bosons, which simultaneously 
satisfies Eq.~\ref{diagonalization} and \ref{normalization}; this 
can be done with the help of generalized matrix diagonalization. 
Unlike fermionic diagonalization, the existence of such a solution
is not guaranteed, unless spin segments are decoupled, {\it i.e.}
$J_{AF} = 0$.
Such unfortunate situations are, in general, associated with modes of
zero energy, which then leads to divergent number of bosons (or
divergent magnetization). 
In the case of a single ferromagnetic chain,
Takahashi~\cite{takahashi86} introduced a chemical potential (or a
magnetic field) to overcome the difficulty. 
In this case, we need to introduce to the Hamiltonian a set of $M_s$
local chemical potentials, one for each segment, to find a solution. 
These chemical potentials are chosen such that the magnetization of each
spin segment (equivalently, of each spin with periodic boundary
condition) is zero.
These constraints, which have the physical significance of restoring the
rotational symmetry, ensure that the number of bosons be finite at
finite temperatures.  
The thermodynamic quantities can be calculated from the excitation
energies $\varepsilon$'s and the transformation matrix $V$. 
In particular, magnetic susceptibility per site $\chi$ can be written as 
\begin{equation}
\chi = {1 \over 3TN} \sum_{s = 1}^{M_s} \sum_{i = 1}^{N_s} 
\tilde{n}_i^s \left ( \tilde{n}_i^s + 1 \right ),
\end{equation}
where $\tilde{n}_i^s$ is the occupation number for the boson correspond
to operator $\alpha_i^s$ and excitation energy $\varepsilon_i^s$,
\begin{equation}
\tilde{n}_i^s = \langle \alpha_i^{s\dagger} \alpha_i^s \rangle 
= {1 \over e^{\varepsilon_i^s / T} - 1}.
\end{equation}

\section{Results for magnetic susceptibility}
\label{sec:results}

\begin{figure}
{\centering \includegraphics{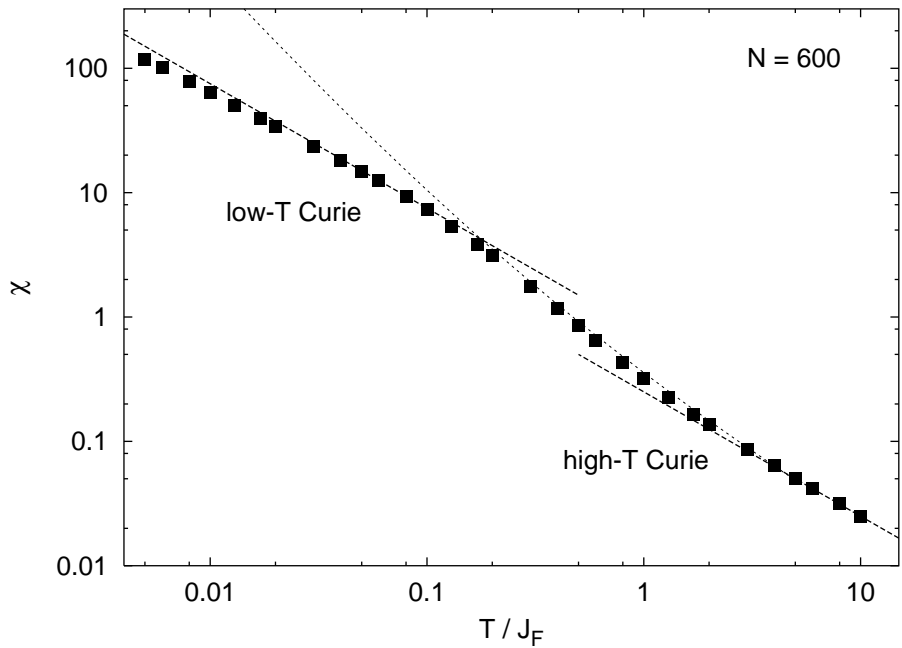} }
\caption{\label{fig1}
Static magnetic susceptibility per spin ($\chi$) of a single random
Heisenberg spin chain of 600 spins. 
The spins are coupled with their nearest neighbors by ferromagnetic
couplings ($J_F$), with 59 antiferromagnetic impurity couplings ($J_{AF}
= 0.5 J_F$), so that the spin chain can be viewed as 60 ferromagnetic
segments of average length 10 (spins).  
On the log-log scale, the two dashed lines with slope unity are exact
results of the ordinary Curie susceptibility of independent spins at high
temperatures (Eq.~\ref{highTexpansion}) and the low-temperature Curie
susceptibility expected by the RSRG approach (Eq.~\ref{rsrg}). 
The dotted line is the low temperature expansion of the magnetic
susceptibility (Eq.~\ref{lowTexpansion}) of a pure spin-1/2 ferromagnetic
chain (of infinite length), which is proportional to $1/T^2$ in the low
temperature limit.  
}
\end{figure}

Figure~\ref{fig1} shows the static magnetic
susceptibility $\chi$ per spin of a 600-spin ferromagnetic chain, 
with 59 antiferromagnetic impurity bonds, 
which divide the chain into 60 ferromagnetic segments, 
{\it i.e.}, $p = 0.1$. 
The impurity bonds ($ J_{AF} = 0.5$) are put randomly with the
constraint that the total magnetization of the corresponding classical
ground state is zero. 
This guarantees that $ \chi $ drops to zero at zero temperature.
The susceptibility curve shows three temperature regimes. 
For $T / J_F > 3$, result can be fit to a high-temperature Curie law
(Eq.~\ref{highTexpansion}), suggesting that spins behave independently.
For $0.2 < T / J_F < 3$, $\chi$ rises with decreasing temperature,
following Eq.~\ref{lowTexpansion}. 
This implies that spins start to correlate, forming independent
ferromagnetic segments. 
The antiferromagnetic coupling is still too weak to affect
thermodynamics of the random spin chain at these temperatures. 
Below $T = 0.2 J_F$, $\chi$ is in good agreement with 
\begin{equation}
\label{lowTCurie}
\chi = {c_{LT} \over T}, \hspace{1cm} c_{LT} = {3 \over 4},
\end{equation}
the low-temperature Curie behavior expected by the RSRG approach
(Eq.~\ref{rsrg} for $p = 0.1$).  
We point out that the Curie constant of independent block
spins (Eq.~\ref{independentSegments}) is expected to be $c_{BS} = 7/4$
for $p = 0.1$. 
Therefore, the low-temperature Curie constant is indeed reduced by
antiferromagnetic couplings between ferromagnetic segments.   
We note that $\chi$ deviates from Eq.~\ref{lowTCurie} below $T = 0.02
J_F$.    
We believe this is an artifact due to finite size (we have only 600
spins), since there are no more segments to decimate. 
$\chi$ bends down as we specially require that $\chi$ go to zero at
zero temperature.  
Since the number of iterations required to find a solution for the
generalized Bogoliubov transformation grows significantly at low
temperatures (nearly 1000 iterations at $T = 0.005$), we were unable to
approach temperatures much lower than $T = 0.01$ or to calculate many
samples on our regular workstations.  
It is worth pointing out that we can identify the crossover temperature
from the ferromagnetic spin-chain physics to random spin-chain physics
as the cross point of Eq.~\ref{lowTCurie} and \ref{lowTexpansion}. 
In the dilute doping limit (small $p$), we have 
\begin{equation}
T_x = J_F{p \over 1-p} \left [ 0.50 + 3.04 \sqrt{p \over 1-p} 
+ O\left( {p \over 1-p} \right ) \right ]
\end{equation}
For $p = 0.1$, $T_x = 0.168 J_F$. 

\begin{figure}
{\centering \includegraphics{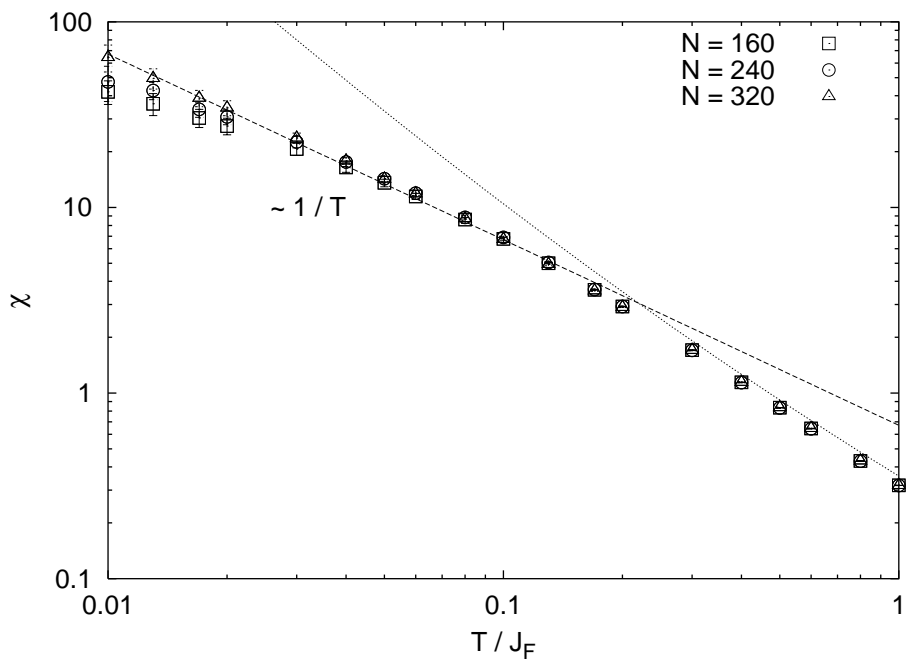} }
\caption{\label{fig2}
Sample averaged magnetic susceptibility per spin for spin chain of 160,
240, 320 spins, which consists of 20, 30, 40 ferromagnetic segments,
respectively, coupled by impurity antiferromagnetic bonds with 
$ J_{AF}=0.5 J_F $. 
The dotted line is the low temperature expansion of the magnetic
susceptibility (Eq.~\ref{lowTexpansion}) of a spin-1/2 ferromagnetic
chain (of infinite length).
The dashed line is a fit to the apparent (see text for detail)
low-temperature Curie susceptibility. 
}
\end{figure}

Figure~\ref{fig2} shows the susceptibility per spin 
$ \chi $ of spin chains of 20, 30, and 40 ferromagnetic segments 
({\it i.e.}, with 19, 29, and 39 antiferromagnetic impurity bonds,
respectively).  
The average ferromagnetic segment length is fixed at 8 spins ($p =
1/8$), so that the lengths of the corresponding spin chains are 160,
240, and 320, respectively.  
The antiferromagnetic coupling $ J_{AF} $ is chosen again to be 0.5.
$ \chi $ is averaged over 10-20 random realizations depending on size. 
For $ 0.2 < T/J_F < 1 $, $ \chi $ roughly follows the low-temperature
(compared with $ J_F $) expansion of a single ferromagnetic chain
(Eq.~\ref{lowTexpansion}). 
From now on, we neglect the thermodynamics at $T > J_F$ (although it is
capable of being calculated within our theory), since it is trivial and
not the interest of this paper.  
Below the crossover temperature $ T_x \sim 0.2 $, $ \chi $ deviates from
the pure chain susceptibility, and obeys a Curie law. 
The range of the Curie susceptibility extends to lower and lower
temperature when we have more segments (from 20 to 40 in
Fig.~\ref{fig2}), therefore longer chain, for a
fixed impurity concentration.  
At $ N_s = 40 $, this range extends beyond one decade for $p = 1/8$. 
We point out that the Curie constant obtained from
Fig.~\ref{fig2} ($c = 0.67$) is roughly 15\% larger than
predicted by Eq.~\ref{rsrg}. 
This should be viewed as a mixture of both the RSRG-predicted behavior
and the ferromagnetic-chain physics. 
The reason will become clear later after we explore the impurity
concentration dependence of the Curie behavior. 

\begin{figure}
{\centering \includegraphics{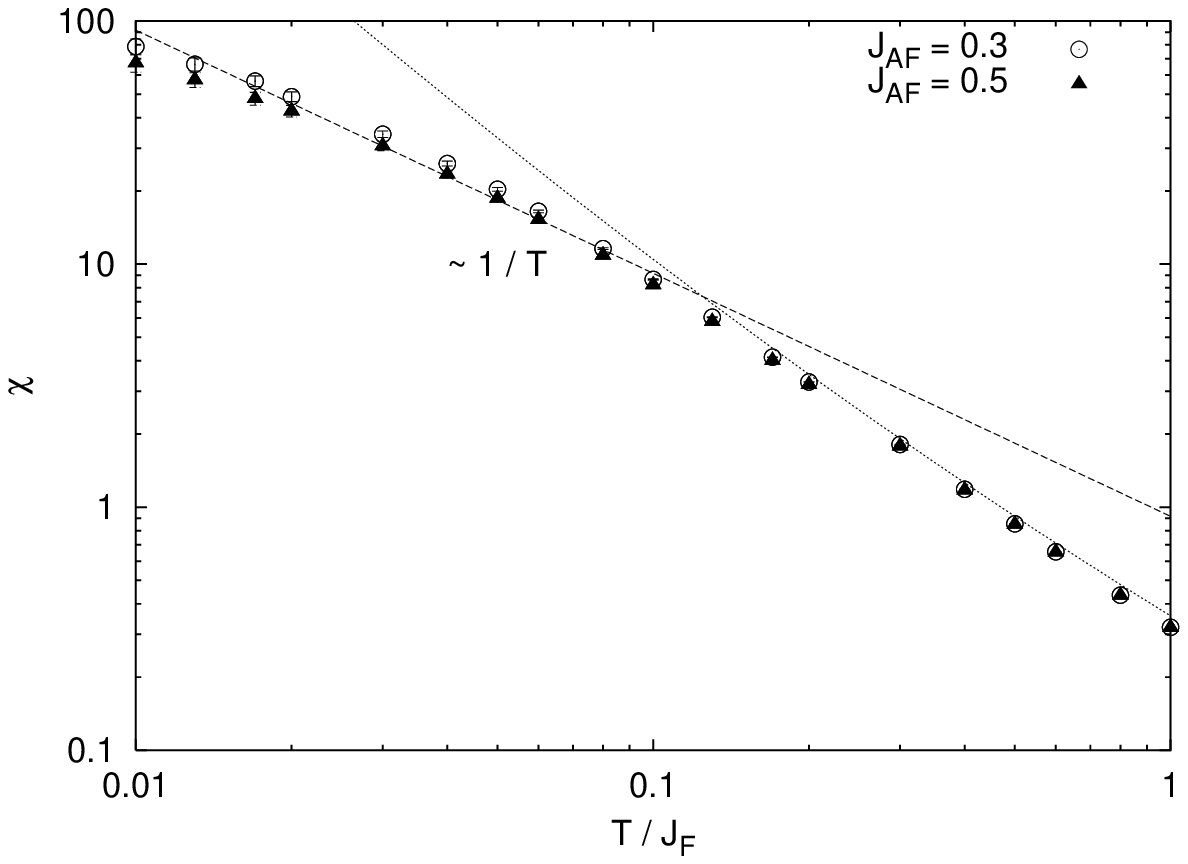} }
\caption{\label{fig3}
Sample averaged magnetic susceptibility per spin for spin chains of 360
spins in 30 ferromagnetic segments, coupled by antiferromagnetic
bonds with $ J_{AF}=0.3 J_F $ and $ 0.5 J_F $. 
The data is expected to agree with the low-temperature Curie
susceptibility (the dashed line) in the RSRG method. 
The temperature range of the agreement is expected to be higher for
stronger antiferromagnetic impurity bonds ($ J_{AF}=0.5 J_F $).  
The dotted line is the low temperature expansion of the magnetic
susceptibility (Eq.~\ref{lowTexpansion}) of a spin-1/2 ferromagnetic
chain (of infinite length).
}
\end{figure}

The antiferromagnetic coupling strength $ J_{AF} $ determines the
temperature scale at which the ferromagnetic segments couple to their
nearest neighbors. 
The smaller $ J_{AF} $ is, the lower the temperature at which RSRG results are
valid becomes, leading therefore to an increasingly wider crossover from
the physics of a pure ferromagnetic chain.  
Figure~\ref{fig3} shows the sample averaged magnetic
susceptibility per spin $\chi$ for $ J_{AF} = 0.3 $ and 0.5 for spin
chains of 30 ferromagnetic segments. 
The average number of spins in each segment is 12 spins ($p = 1/12$). 
The dashed line corresponds to a Curie constant $c = 11/12$, expected by
Eq.~\ref{rsrg} for $p = 1/12$. 
Although the system size (360 spins) is probably not large enough, one
can nevertheless identify the Curie behavior for $J_{AF} = 0.5$ around 
$T = 0.05$ and for $J_{AF} = 0.3$ around $T = 0.02$. 
Note that $\chi$ for $J_{AF} = 0.3$ is greater than that for $J_{AF} =
0.5$ at low temperatures. 
This is expected because decimated spin segments, in the RSRG language,
contribute less to the magnetic susceptibility than the sum of original
segments, and because the decimation happens at lower temperatures for
smaller $J_{AF}$.  
For very small $J_{AF}$, we expect the system crosses over from a
ferromagnetic regime at high temperatures, first, to an (almost)
independent block-spin regime at lower temperatures, characterized by a
Curie constant as in Eq.~\ref{independentSegments}.
There is another crossover, at very low temperatures (dependent on
$J_{AF}$), from the independent block-spin regime to the RSRG regime,
characterized by a smaller Curie constant as in Eq.~\ref{rsrg}.

\begin{figure}
{\centering \includegraphics{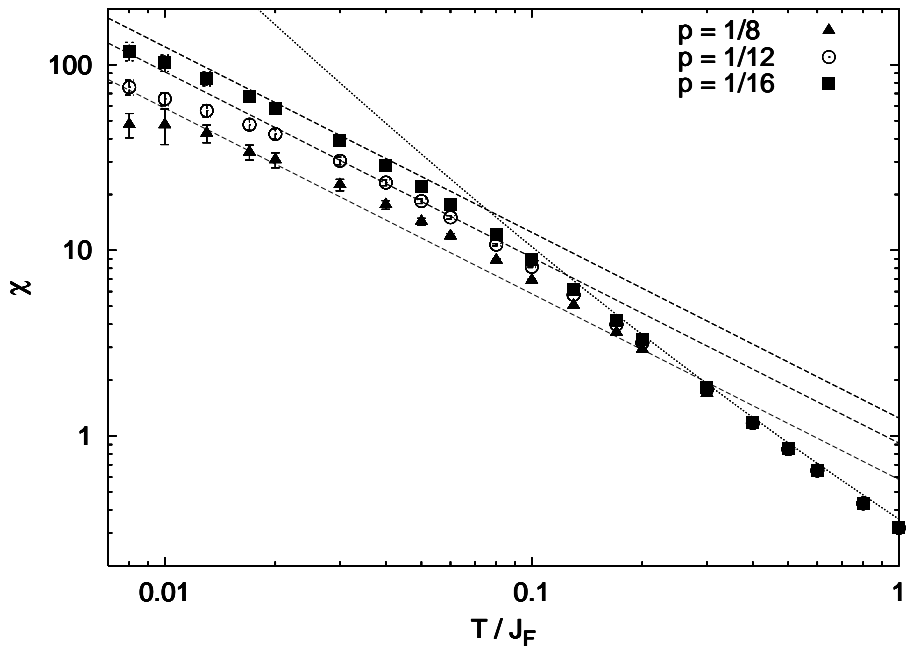} }
\caption{\label{fig4}
Sample averaged magnetic susceptibility per spin for spin chains of 240,
360, and 480 spins, each containing 30 ferromagnetic segments, coupled
by antiferromagnetic bonds with $ J_{AF}=0.5 J_F $. 
The dotted line is the low temperature expansion of the magnetic
susceptibility (Eq.~\ref{lowTexpansion}) of a spin-1/2 ferromagnetic
chain (of infinite length).
The dashed lines are the exact results of the low-temperature Curie
susceptibility expected by the RSRG approach for the three impurity
concentrations.
}
\end{figure}

We have also studied the dependence of the magnetic susceptibility on
the impurity concentration, or equivalently, average number of spins in
each segment. 
Qualitatively, the larger number of spins within each segment increases
the magnetic susceptibility per spin, since $\chi \sim N_s^2$ for a
ferromagnetic $N_s$-spin segment (for large $N_s$).
As a consequence, the Curie constant is expected to increases with the
average number of spins in each segment, or more precisely, as in
Eq~\ref{rsrg}.  
This trend is shown in Fig.~\ref{fig4}, where we
plot the averaged susceptibility per spin for 30-segment spin chains,
with 8, 12, and 16 spins per segment on average, respectively.
The three dashed lines are the low-T Curie behavior the RSRG approach
expected (Eq.~\ref{rsrg}).
As is pointed out in earlier paragraph, our data is in significant
disagreement with Eq.~\ref{rsrg} for $p = 1/8$. 
For $p = 1/16$, one finds that $\chi$ follows Eq.~\ref{rsrg} from $T =
0.015 J_F$ to $0.04$, the regime antiferromagnetic impurity bonds are
dominating.  
The Curie regime shifts to $ 0.02 < T/ J_F < 0.08 $ for $p = 1/12$. 
This is consistent with Eq.~\ref{effectiveJ} that longer segments have
relatively weaker effective intersegment coupling.
The trend suggests that for $p = 1/8$ the Curie regime would occur above
$T > 0.1 J_F$, which coincides with the regime where intersegment
coupling is important. 
Therefore, we believe the discrepancy between our data and
Eq.~\ref{rsrg} reveals a collective contribution from the ferromagnetism
with each segment and the antiferromagnetism between segments. 
This demonstrates that inappropriate parameters can kill (or lead to the
wrong) RSRG physics.
We note that the crossover regime (deviating from both
Eq.~\ref{highTexpansion} and the $1/T$-law) becomes wider for smaller
$p$. 
This is again consistent with the translation between low concentration
and weak effective intersegment coupling, which, we have demonstrated,
leads to wider crossover regime.

\section{Conclusion}
\label{sec:conclusion}

In the study of the random AFM-FM (nearly ferromagnetic) spin-1/2 chain, 
the modified spin-wave theory allows us to obtain the thermodynamic
properties over a broad temperature range (on logarithmic scale). 
We have studied systems as large as 600 spins, with up to 60
ferromagnetic segments, for temperature ranges of over three orders of
magnitude. 
At very high temperatures, the spins are independent;
correlation starts to develop within each
ferromagnetic segments as temperature drops, giving rise to
characteristics of a pure ferromagnetic chain.
At low temperatures, the observation of the Curie-like behavior reveals
the regime where ferromagnetic segments are correlated into clusters,
whose size grows with decreasing temperature.  
Therefore, we can study, fairly systematically, in the modified
spin-wave theory the crossover between the high-temperature Curie 
behavior (independent spins) and low-temperature Curie behavior
(correlated spin segments). 

System size is important in observing the low-temperature Curie
behavior, since it determines whether the RSRG scaling regime can be
reached. 
Our study has found that the modified spin-wave theory gives results
in accordance with the RSRG argument for chains of several hundred
spins (after averaging over random bond distribution). 
The study of the effects of impurity bond strength shows
that the low-temperature Curie regime can be consistently pushed
to lower temperatures by decreasing the antiferromagnetic impurity bond
strength.

In the dilute doping regime, we have studied the effects of impurity
concentration on the low-temperature thermodynamics. 
The results are again in fairly good agreement with the RSRG
prediction. 
The exception happens at higher concentration ($p = 1/8$), where we
believe the intrasegment ferromagnetic coupling and the intersegment
antiferromagnetic coupling take actions at about the same temperature
regime, leading to a mixed behavior which appears to look like a
different Curie behavior. 
Further explorations using other methods would be very useful to confirm
our explanation. 

In our study, both weak impurity bonds and low impurity concentration
give rise to a wide crossover from the ferromagnetic spin-chain regime
to the random spin-chain regime. 
Although the modified spin-wave theory limits us to only a portion of
the whole random spin-chain space, we believe such wide crossover can
be generic and it can prevent the real RSRG behavior from being observed
at reasonably low temperatures. 
In fact, the RSRG study of random spin chains~\cite{westerberg97} has
pointed out the possibility of having a crossover region of more than
five orders of magnitude before the true RSRG scaling regime can be
reached.
A quantum Monte Carlo study of the random AFM-FM spin-1/2
chain~\cite{ammon99} has confirmed that a region of weakly interacting
spin segments can exist, whose crossover to the low-temperature scaling
regime is marked by a (somewhat small) peak in the specific heat. 

The main difficulty of directly comparing our results with experiments
seem to be that quasi-1D Heisenberg ferromagnetic systems are
rare since there usually exist interchain couplings which lead to
three-dimensional (3D) ordering at low temperatures. 
For Sr$_3$CuIrO$_6$, long-range (presumably 3D) order develops at 20.1
K without signature of 1D magnetism.~\cite{irons00}  
Recently, however, effects of magnetic impurities on the quasi-1D
ferromagnetic spin chain have been reported for organic radical alloy,
($p$-CDpOV)$_{1-x}$($p$-BDpOV)$_x$,~\cite{mukai99} whose
interchain-intrachain coupling ratio is as small as $J'/J = 3.7 \times
10^{-3}$. 
This may open a door where the low-temperature Curie susceptibility of
the RSRG origin can be observed in ferromagnetic bond dominated random
spin chains.

\section{Acknowledgments}

This work was supported by NSF grants No. DMR-9971541 (X.W. and K.Y.),
No. DMR-9809483 (R.N.B.), the state of Florida (X.W.), and the Alfred P.
Sloan Foundation (K.Y.)

\begin{appendix}

\section{Modified spin-wave theory for ferromagnetic spin segments}
\label{singleSegment}

In Takahashi's original paper~\cite{takahashi86}, the modified spin-wave
theory is aimed at ferromagnetic spin chains with infinite length. 
Nevertheless, the theory can be easily applied to a finite-size
ferromagnetic spin chain with periodic boundary conditions. 
This only changes the continuous spin-wave spectrum to a discrete
spectrum. 
Therefore, for an $N$-spin ferromagnetic chain, we can write the
self-consistent equations for the single (global) chemical potential
$\mu$:
\begin{eqnarray}
\label{selfconsistent}
NS &=& \sum_k \tilde{n}_k, \nonumber \\
\tilde{n}_k &=& {1 \over e^{[\epsilon (k) - \mu]/T} - 1},
\end{eqnarray}
where the discrete spin-wave energy spectrum is
\begin{equation}
\epsilon (k) = 2JS (1 - \cos ka), 
\hspace{1cm} k = {2 \pi i \over Na}, \ i = 0, 1, ..., N-1.
\end{equation} 
The magnetic susceptibility per spin can be expressed as
\begin{equation}
\chi = {1 \over 3TN} \sum_k \tilde{n}_k (\tilde{n}_k + 1).
\end{equation}

Analytical results, which turn out to be exact, can be obtained for both
high and low temperature limit. 
In the high temperature limit, $\tilde{n}_k = S$ is a constant. 
Self-consistent Eqs.~\ref{selfconsistent} give 
\begin{equation}
\mu = T \ln \left (1 + S^{-1} \right ).
\end{equation}
Therefore, we have
\begin{equation}
\chi = {S(S+1) \over 3T},
\end{equation}
which is the exact result for independent spins. 
In the low temperature limit, only $k = 0$ mode can be excited,
{\it i.e.} $\tilde{n}_k = NS \delta_{k,0}$. 
Self-consistent Eqs.~\ref{selfconsistent} give 
\begin{equation}
\mu = T \ln \left (1 + N^{-1}S^{-1} \right ).
\end{equation}
Therefore, we have
\begin{equation}
\chi = {S(NS+1) \over 3T},
\end{equation}
which is the exact result for the ferromagnetic ground state with
maximum spin.   

\begin{figure}
{\centering \includegraphics{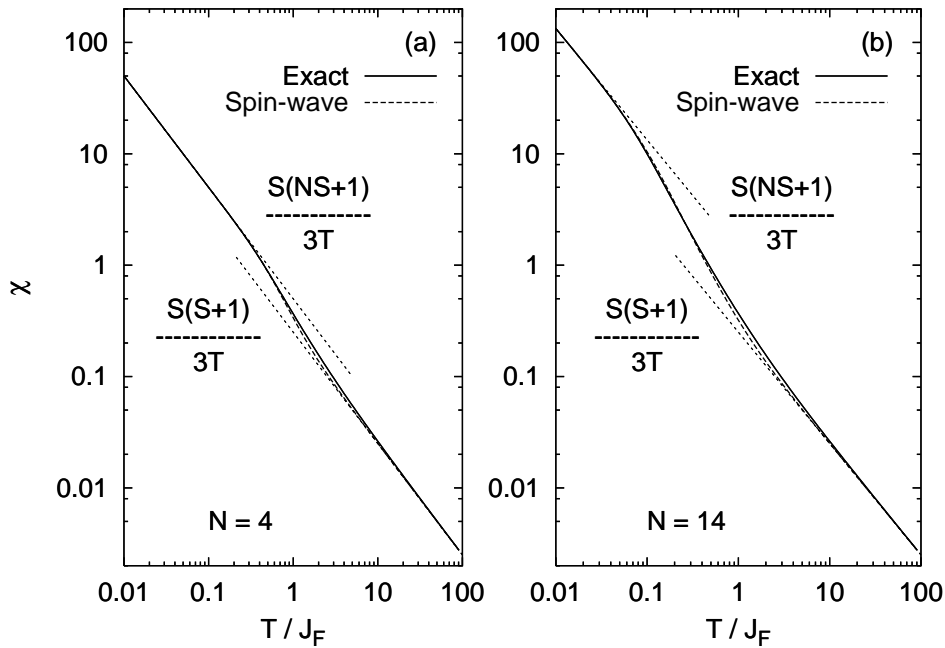} }
\caption{\label{fig5}
Exact diagonalization result (solid line) and the modified spin-wave
theory result (dashed line) of the magnetic susceptibility per spin 
for (a) $N = 4$ and (b) $N = 14$ ferromagnetic spin segments (S = 1/2)
with periodic boundary conditions. 
The two dotted lines are the high-temperature Curie susceptibility of
independent spins and the low-temperature Curie susceptibility of the 
spins when frozen into a block spin. 
}
\end{figure}

Figure~\ref{fig5} shows $\chi$ for ferromagnetic
spin-1/2 chains of finite length $N = 4$ and 14. 
The modified spin-wave theory results agree very well with exact results
for both sizes. 
The largest descrepency (less than 15\%) occurs around $T = J_F$, at
which independent spins start to correlate.
This regime is of less interest for the purpose of this paper.
In both the high temperature limit ($T > 10 J_F$) and the low
temperature limit ($T < \Delta_{SW}$), the modified spin-wave theory
gives exact results. 
This is a very impressive result since the spin-wave theory is in
principle developed around $T = 0$, describing the low temperature
excitations of the ferromagnetic spin chain. 
In addition, we are able to show that boundary conditions have only weak
effects on the rapidly changing susceptibility.
In particular, behavior in the high temperature limit and the low
temperature limit remains unchanged. 

\section{Coupling between two ferromagnetic spin segments}
\label{twoSegments}

\begin{figure}
{\centering \includegraphics{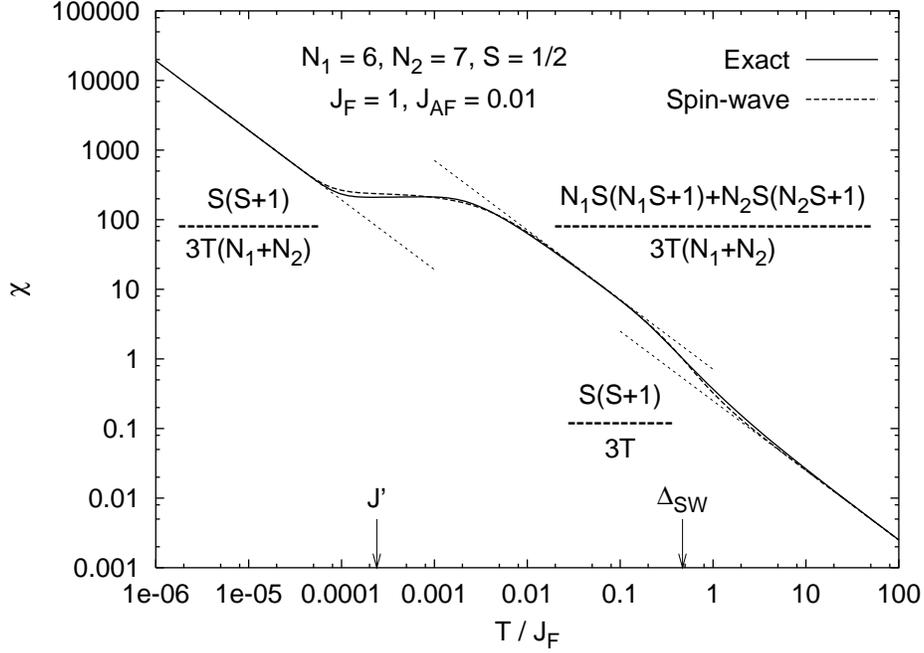} }
\caption{\label{fig6}
Exact diagonalization result (solid line) and the modified spin-wave
theory result (dashed line) of the magnetic susceptibility per spin of
two spin-1/2 ferromagnetic segments ($N_1 = 6$ and $N_2 = 7$) coupled by
a $J_{AF} = 0.01 J_F$ antiferromagnetic bond. 
The three dashed lines describe the Curie behavior in the following
three regimes. 
(1) At high temperatures, spins are independent. 
(2) At intermediate temperatures, spins within each ferromagnetic
segments are locked into a block spin with no internal excitations. 
The two block spins interacts with each other through the effective
antiferromagnetic coupling ($J'$) generated by the antiferromagnetic
bond ($J_{AF}$).  
(3) At low temperatures, the two block spins are locked into one S = 1/2
spin.   
}
\end{figure}

The problem of two ferromagnetic spin segment coupled by an
antiferromagnetic bond is a special case of more general random spin
chains. 
For small systems, we can compare our modified spin-wave
theory with exact diagonalization. 
Figure~\ref{fig6} shows the susceptibility for two
ferromagnetic segments ($J_F = 1$) coupled with a $J_{AF} = 0.01$
antiferromagnetic bond.  
The numbers of spins in the two segments are $N_1 = 6$ and $N_2 = 7$. 
Three regimes where the susceptibility obeys $1/T$-law are well-defined
in Fig.~\ref{fig6}. 
The corresponding pictures from high to low temperature are independent
spins, two independent spin segments, completely locked spins,
respectively. 
The susceptibility can be fit very well to the Curie law obtained from
the simple pictures. 
In these Curie regimes, the modified spin-wave theory gives exact
results. 
Discrepancies between the modified spin-wave results and exact results
can only been noticed at the two crossover regimes between them. 
Crossover temperature scales can be roughly estimated by the smaller
spin-wave gap $\Delta_{SW}$ of the two segments and the
effective antiferromagnetic coupling $J'$ between the two segments, as
marked by arrows in Fig.~\ref{fig6}. 

\begin{figure}
{\centering \includegraphics{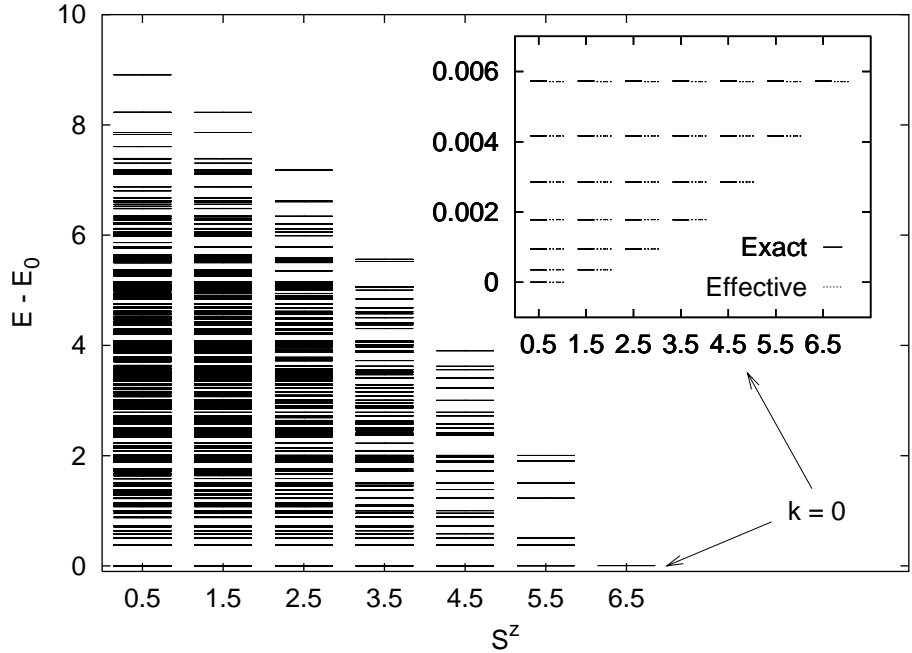} }
\caption{\label{fig7}
Energy spectrum of the two ferromagnetic segments described in
Fig.~\ref{fig6} ($N_1 = 6$ and $N_2 = 7$ spin segments
coupled by a $J_{AF} = 0.01$ antiferromagnetic bond).
The lowest energy manifold ($k = 0$ modes) of the exact diagonalization
result (solid lines), shown on an amplified scale in the inset, is
compared with the energy spectrum of the two block spins coupled by the
effective coupling $J'$ (Eq.~\ref{effectiveJ}).
}
\end{figure}

In the RSRG scheme, one introduces an effective antiferromagnetic
coupling $J'$ between the two segments, which replaces the coupling
between the two neighboring end spins. 
By inspecting the energy spectrum of the two-segment system in
Fig.~\ref{fig7}, one can explore the validity of the replacement. 
At low enough temperatures ($T < \Delta_{SW}$), the finite spin-wave
excitations are frozen within each segment. 
Therefore, only $k = 0$ modes are relevant when the two segments
couple with the effective $J'$.  
This results a group of lowest energy levels ($S = 1/2$, 3/2, ...,
13/2), which are separated from 
higher energy levels by a energy gap ($\sim \Delta_{SW}$). 
The gap is expected to become comparable with the width of the $k = 0$
energy levels, when $J_{AF}$ is close to $J_F$, when RSRG argument
becomes less accurate. 
For small $J_{AF}$, these $k = 0$ modes obtained from the exact
diagonalization are in good agreement with the effective levels in the
RG picture (the inset of Fig.~\ref{fig7}).
The difference in energy levels given by the two methods grows with
increasing $J_{AF}$. 
Nevertheless, the agreement in the susceptibility obtained
from the two methods still persists even when $J_{AF} \sim J_F$, as long
as $T \ll \Delta_{SW}$, which poses a temperature restriction on
the RSRG scheme. 

\end{appendix}


\end{document}